\documentclass{jfm}
\usepackage{graphicx}
\usepackage{epstopdf,epsfig}
\usepackage{amsmath}

\usepackage{tikz}
\usetikzlibrary{arrows}

\newcommand{\ee}{{\mathrm e}}
\newcommand{\ii}{{\mathrm i}}
\newcommand{\dd}{{\mathrm d}}

\usepackage{url}

\shorttitle{A bulk-interface correspondence for equatorial waves}
\shortauthor{C. Tauber, P. Delplace,  and A. Venaille}

\title{\textcolor{black}{A bulk-interface correspondence for equatorial waves}}

\author{C. Tauber\aff{1},
  P. Delplace \aff{2}
 \and A. Venaille \aff{2}}

\affiliation{\aff{1} Institute for Theoretical Physics, ETH Z\"urich, Wolfgang-Pauli-Str. 27, CH-8093 Z\"urich, Switzerland
\aff{2}  Univ Lyon, Ens de Lyon, Univ Claude Bernard, CNRS, Laboratoire de Physique, F-69342 Lyon, France}

\begin{document}

\maketitle

\begin{abstract}
Topology is bringing new tools for the study of fluid waves. The existence of unidirectional Yanai and Kelvin equatorial waves has been related to a topological invariant, the Chern number, that describes the winding of $f$-plane shallow water eigenmodes around band crossing points in parameter space. In this previous study, the topological invariant was a property of the interface between two hemispheres. Here we ask whether a topological index can be assigned to each hemisphere. We show that this can be done if the shallow water model in $f$-plane geometry is regularized by an additional odd-viscous term. We then compute the spectrum of a shallow water model with a sharp equator separating two flat hemispheres, and recover the Kelvin and Yanai waves as two exponentially trapped waves along the equator, with all the other modes delocalized into the bulk. \textcolor{black}{This model provides an exactly solvable example of bulk-interface correspondence in a flow with a sharp interface, and offers a topological interpretation for some of the transition modes described by [Iga, Journal of Fluid Mechanics 1995].} It also paves the way towards a topological interpretation of coastal Kelvin waves along a boundary, and more generally, to an understanding of bulk-boundary correspondence in continuous media.
%Beyond the shallow water case, our study shows that geophysical provide non-conventional manifestations of bulk-edge correspondence,
\end{abstract}


\section{Introduction}

Tools from topology developed over the last decades in condensed matter physics have recently shed new light on our understanding of fluid waves, from the design of microfluidic devices \citep{souslov2017topological}, to acoustic waves \citep{yang2015topological}, to planetary atmospheres \citep{delplace2017topological,perrot2018topological}, as well as in active matter flows \citep{shankar2017topological}. It has been realized that important and robust information on the spectrum of a linear operator is encoded into the eigenmodes of this operator in unbounded geometry, with constant coefficients. This information is revealed by the winding of the eigenmodes parameterized over a closed surface. This winding is a topological invariant, the Chern number, that can be explicitly computed. For instance, \cite{delplace2017topological} showed that inertia-gravity waves in the rotating shallow water model exhibit such a topological property in $(k_x,k_y,f)$-space, with $f$ the Coriolis parameter, and  $(k_x,k_y)$ the wavenumber. More precisely, a Chern number of $2$ was found for the positive frequency inertia-gravity waves as they enclose the origin in parameter space, while the zero-frequency (geostrophic) modes carry a vanishing Chern number. \\ 
\indent One spectacular manifestation of these singularities occurs when one computes the spectrum of the same operator, now assuming that the one of the parameters varies spatially. In the shallow water model, such computation have for instance been performed by \cite{matsuno1966quasi} on the equatorial beta plane, assuming linear variations of the Coriolis parameter in the $y$ (meridional) direction. He discovered the existence of two branches in the dispersion relation that transit between different wave bands when the wavenumber in the $x$ (zonal) direction is varied: the equatorial Kelvin wave, and the mixed Rossby-gravity wave, now known as the Yanai wave. These modes are more localized along the equator than the others, and are unidirectional. The gradient of planetary vorticity also supports the propagation of low frequency planetary Rossby waves, that lift part of the degeneracy of geostrophic modes. However, contrary to the Kelvin and the Yanai waves, the Rossby waves remain in the geostrophic band. Thus, in the equatorial beta plane, when the zonal wavenumber is varied from negative to positive values,  the positive frequency inertia-gravity wave band has a net gain of two modes \citep{delplace2017topological}. The correspondence between a topological invariant, the first Chern number, that describes  degeneracy points for bulk waves in parameter space, on the one hand, and the number of modes that transit from one band to another in the equatorial beta-plane on the other hand, is reminiscent of the Atiyah-Singer index theorem \citep{faure2018expose,bal2018continuous} and could be referred to as a \textit{topological spectral flow correspondence}. This correspondence has been proven useful to interpret molecular spectra \citep{faure2000topological}, or Lamb-like waves trapped along an interface of density stratification \citep{perrot2018topological}, among other applications in physics \citep{nakahara2003geometry}.

\indent There exists, however, the possibility for a stronger \textit{bulk-interface correspondence},  when a topological index can be assigned to each wave band on each side of the interface, rather than to a band-crossing point in parameter space.  The bulk-interface correspondence then predicts the number of unidirectional edge states trapped along the interface between the two regions as a consequence of the mismatch between the topological indices of two wavebands across the interface. In condensed matter, thanks to an underlying lattice structure, this bulk-interface correspondence naturally follows from a \textit{bulk-boundary correspondence} that relates a topological index of the bulk with the number of uni-directional edge modes propagating along a boundary \citep{hatsugai1993chern,graf2013bulk}. In this context the interface is recovered by gluing together the boundaries of two topologically distinct systems. It is thus natural to ask whether  equatorial Kelvin and Yanai waves can also be understood as topological interface states between two distinct topological systems. In other words, can a single hemisphere  be interpreted as a topological media on its own? Here we use odd viscous terms to assign a well-defined topological invariant to each hemisphere, building on previous work by \cite{volovik1988analogue,souslov2018topological}, and show that a bulk-interface correspondence is satisfied at the equator. 

This example of the linearized rotating shallow water model constitutes {\color{black} an exactly solvable and physically relevant example}  well suited to clarify  ongoing issues related to bulk-interface correspondence in continuous media \citep{faure2018expose,bal2018continuous}\textcolor{black}{, in a case where the interface is sharp}.  It also offers a novel interpretation of the equatorial waves  as two edge states propagating along a sharp equator separating two flat hemispheres that are topologically distinct, with an explicit computation of the spectrum that is complementary to the beta plane case considered by \cite{matsuno1966quasi}\textcolor{black}{, and to the interpretation in terms of transition modes \citep{iga1995transition}, who used arguments based on the conservation of zeros in eigenfunctions.} 
 
\textcolor{black}{Finally, this case of a \textit{sharp equator} for the Coriolis parameter  provides a first step  towards an understanding of the more complicated case of a \textit{sharp boundary} for the flow domain, that includes the case of coastal Kelvin waves. In the presence of a boundary, \cite{iga1995transition} found that coastal Kelvin waves can be removed from the spectrum just by changing the boundary conditions along the wall. This seems to contradict the expected topological robustness of the boundary modes.  It is then natural to ask whether this conclusion is robust to the addition of odd-viscosity \citep{souslov2018topological}. We will indeed confirm Iga's result in that case, which raises an apparent paradox on the applicability of bulk-edge correspondence in fluids.}

\section{Linearized rotating shallow water equations with odd viscosity}

We consider the rotating shallow water equations linearized around a state of rest in Cartesian geometry, with an additional \textit{odd viscosity} term of amplitude $\epsilon$ \citep{avron1998odd}
\begin{align}
\partial_t \eta &= - \partial_x u - \partial_y v\\
\partial_t u &=-\partial_x \eta+ \left(f +\epsilon  \nabla^2 \right)v \\
\partial_t v &=-\partial_y \eta- \left(f +\epsilon  \nabla^2 \right)u 
\end{align}
where $(u,v)$ are the two (depth independent) velocity components of the flow in the plane $(x,y)$, $\eta$ the interface elevation relative to the mean depth $H$, $f$ the Coriolis parameter that may depends on spatial coordinates. The horizontal variations of interface height corresponds to pressure gradients, as the thin layer of fluid satisfies hydrostatic balance. Time unit have been chosen such that the shallow water phase speed is $\sqrt{gH}=1$, with $g$ the standard gravity. 

{\color{black} It was realized two decades ago that  a two-dimensional system with broken time reversal symmetry at a microscopic level must include an 'odd viscosity' term \cite{avron1998odd}. This term has the same effect on the flow evolution as the Coriolis force, except that its strength depends on the wavenumber. It may be thought of as the first order correction (in $k$) to the inviscid dynamics that (i) breaks time reversal symmetry  (ii) preserves isotropy (iii) does not dissipate energy.  The existence of odd-viscous terms is prevented by Onsager reciprocity relations only when the microscopic dynamics is time reversible. This is the case for most classical fluids.  However, rotating flows are not time-reversal symmetric.  Thus, if subgrid-scales dynamics of a rotating systems are modeled by analogy with molecular effects, an odd viscosity term must be included. Such odd viscosity are actually reminiscent of skew diffusion operators, that have already been proven very useful to model the effect of baroclinic instability in coarse resolution oceans models \citep{vallis2017atmospheric}.}

Here, the odd-viscosity term will be essential to assign a topological invariant  to the flow model when the Coriolis parameter $f$ is prescribed, by regularizing pathological features in bundles of eigenmodes at large wavenumbers.  \textcolor{black}{The addition of such terms was also proposed recently by \cite{wiegmann2013hydrodynamics} to describe a fluid of point vortices and by \cite{banerjee2017odd,souslov2018topological} for a flow model similar to shallow water equations, motivated by plasma and active matter applications. In the latter work, regularization is also discussed. Moreover this regularization procedure is well-known in a condensed matter context and probably goes back to \cite{volovik1988analogue} for the Dirac Hamiltonian in two dimensions (see also \cite{bal2018continuous} for a more general description of this problem). In our case we  consider $\epsilon$ as a (arbitrarily small) constant, and will check that known results are recovered in the the limit $\epsilon \rightarrow 0$.  This contrasts with the effect of usual viscosity in three-dimensional turbulence, with the occurrence of anomalous dissipation in the limit of weak viscosity. }

Another (complementary) way of removing singularities at large wavenumbers is to consider lattice models, such as the discrete models used for numerical simulations \citep{delplace2017topological}. In that case, the admissible wavenumbers are defined on a Torus (called the Brillouin zone in condensed matter), and it is possible to compute the Chern number for the bundle of discretized $f$-plane shallow water eigenmode parameterized on this torus. We leave the study of such discrete models to future work, putting here emphasis on continuous models. 

\section{Bulk Waves in the f-plane}

\begin{figure}
\centering
\includegraphics[scale=0.7]{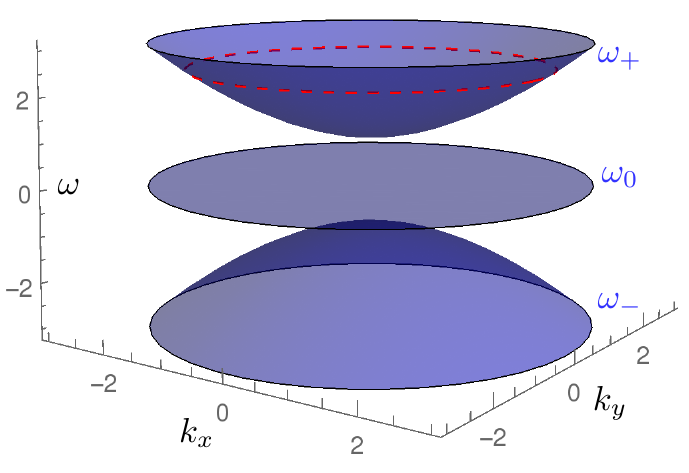}
	%\hspace{0.5cm}
	\includegraphics[scale=0.78]{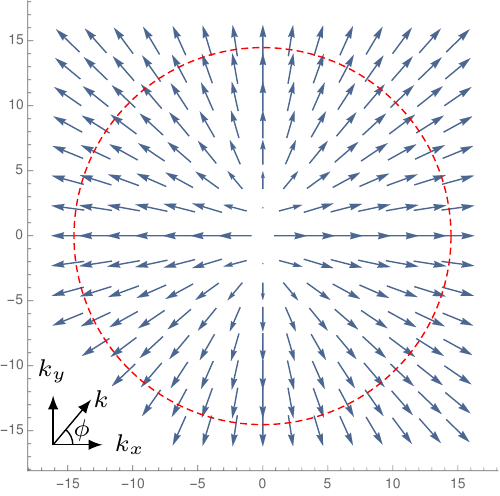}
	%\hspace{0.5cm}
	\includegraphics[scale=0.78]{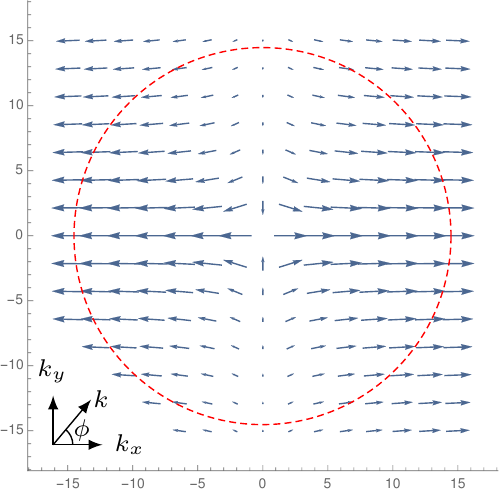}
	\caption{Dispersion relation of the bulk bands $(\omega_-,\omega_0,\omega_+)$ (left) and second component $\hat{u}_{+} \in \mathbb C$ of the positive frequency inertia-gravity eigenmodes $\Psi_+$ in the $(k_x,k_y)$-plane, for $\epsilon  =0.2$ (middle) and $\epsilon =0$ (right). At large $k$ the components $\hat{u}_{+}$ are in both cases multivalued. In the case $\epsilon  \neq 0$ the amplitude of $\hat{u}_{+}$ is constant so that it can become single-valued up to a phase multiplication, allowing for a compactification at $k \rightarrow \infty$. In the case $\epsilon  = 0$ the amplitude at large $k$ is non-uniform so that compactification is impossible. \label{fig:Compactornot}} 
\end{figure}

The \textit{bulk} problem is defined by the case where the flow takes place in an horizontal unbounded plane $(x,y)$ with a given $f$. This is the standard $f$-plane approximation. 
%To simplify the problem we assume $|f \epsilon |\leq 1/4$. 
We look at normal modes of the form $(\eta, u, v) = \ee^{\ii (\omega t -k_x x -k_y y)}(\hat\eta, \hat u, \hat v)$. The previous system becomes
\begin{equation}\label{eq:bulk_hamiltonian}
 \omega \begin{pmatrix}
 \hat \eta \\ \hat u \\ \hat v 
 \end{pmatrix}=
 \begin{pmatrix}
 0 & k_x & k_y \\ k_x & 0 & -\ii (f-\epsilon  k^2) \\ k_y & \ii (f-\epsilon  k^2) & 0
 \end{pmatrix} \begin{pmatrix}
 \hat \eta \\ \hat u \\ \hat v 
 \end{pmatrix}
\end{equation}
where $k^2:= k_x^2+k_y^2$. The band eigenvalues are (see Figure \ref{fig:Compactornot} left)
\begin{equation}\label{eq:omegapm}
\omega_\pm(k_x,k_y) = \pm \sqrt{k^2 + (f-\epsilon  k^2)^2}, \qquad \omega_0(k_x,k_y) = 0
\end{equation}
The middle band $\omega_0$ is flat, and corresponds to odd-geostrophic modes (pressure terms are balanced by Coriolis and odd-viscous terms). The upper band is an interpolation between two parabolas: $\omega_+ \sim |f| + \tfrac{1-2f\epsilon }{2|f|} k^2$ when $k \rightarrow 0$ and $\omega_+ \sim |\epsilon| k^2$ when $k \rightarrow \infty$, and similarly for the lower one. Those modes correspond to odd-Poincar\'e (or odd-inertia gravity) waves. Importantly, those bands are separated by a range of forbidden frequencies as long as $f \neq 0$: the system is \textit{gaped}. As we shall see it is possible to define a topological invariant, the first Chern number, for each band only in presence of odd-viscosity: $\epsilon  \neq 0$.

\subsection{Topology of the upper band}
Each eigenmode of \eqref{eq:bulk_hamiltonian} is defined up to a phase that one can arbitrarily choose locally (gauge freedom), namely for each point of the plane $(k_x, k_y)$. When such modes are defined onto a \textit{closed surface}, the existence of a smooth phase everywhere is not guaranteed. In that case, it is always possible to remove a phase singularity at a given point by a suitable gauge choice, but this singularity has to appear somewhere else over the closed surface. The impossibility to define globally a continuous phase is a topological property of the mode, captured by an integer-valued index, the first Chern number. In our case, the difficulty to study these phase singularities originates from the fact that the parameter space is not a closed surface, but the plane $(k_x,k_y)$. Defining a meaningful topological property for the modes thus requires a \textit{compactification} of the problem. Even though the compactification of a plane to a (Riemann) sphere is a standard procedure, it is not guaranteed that the eigenmodes defined on the plane can be smoothly mapped on the sphere. In particular, we will show that another kind of singularity of the eigenmodes, appearing at infinity in the plane may prevent the compactification, and thus the definition of the topological property. However, this issue can be overcome thanks to the odd-viscosity term.

At this step, it is judicious to switch to cylindrical coordinates $k_x = k  \cos \phi,\,k_y = k \sin \phi$. In particular the bands are $\phi$-invariant. A normalized (hence non-vanishing) family of eigenvectors associated to $\omega_+$ is
\begin{equation}
\Psi_+(k,\phi)  = \dfrac{1}{\sqrt 2} \begin{pmatrix}
k\, /\omega_+(k) \\ \cos \phi - \ii \sin \phi (f-\epsilon  k^2) / \omega_+(k)\\\sin \phi + \ii \cos\phi  (f-\epsilon  k^2)/ \omega_+(k) 
\end{pmatrix}
\end{equation}
up to a phase that can be chosen arbitrarily (gauge freedom). These eigenmodes are \textit{regular} (single-valued) on the punctured plane $(k_x,k_y) \in \mathbb R^2\setminus\{0\}$. We now ask the following questions: 1)\,can we extend the regularity property for $k\rightarrow 0$ and $k \rightarrow  \infty$? 2) If yes, can we do it simultaneously? To answer these questions, first notice that 
\begin{equation}
\lim_{k\rightarrow 0} \Psi_+(k,\phi) = \dfrac{1}{\sqrt 2}\begin{pmatrix}0 \\ \cos \phi - \ii \,\text{sign}(f) \sin \phi \\ \sin \phi + \ii \,\text{sign}(f) \cos \phi  \end{pmatrix} = \ee^{- \ii \,\text{sign}(f) \phi }\dfrac{1}{\sqrt 2}\begin{pmatrix}0 \\ 1 \\ \,\text{sign}(f) \ii  \end{pmatrix}
\end{equation}
where $\,\text{sign}(f) = f/|f|$ is the sign of $f$. $\Psi_+$ is then multivalued, or \emph{singular}, at 0. But this singularity can be removed using gauge freedom. %\footnote{\textcolor{black}{Equivalently, the eigenprojection $P_+(k,\phi) = |\Psi_+\rangle\langle \Psi_+|$ is single-valued as $k\rightarrow 0$.}}. 
We define $\Psi_+^A(k,\phi) := \ee^{\ii \,\text{sign}(f) \phi} \Psi_+(k,\phi)$, implying
%$	\lim_{k\rightarrow 0} \Psi_+^A(k,\phi) = \tfrac{1}{\sqrt 2}(0,1,\,\text{sign}(f) \ii)$
so that $\Psi_+^A$ is single-valued, or \emph{regular} on $\mathbb R^2$. The problem has been \emph{compactified} at $0$. Similarly,
\begin{equation}
\lim_{k\rightarrow \infty} \Psi_+(k,\phi) = \dfrac{1}{\sqrt 2}\begin{pmatrix}0 \\ \cos \phi + \ii \,\text{sign}(\epsilon)\sin \phi \\ \sin \phi - \ii \,\text{sign}(\epsilon)\cos \phi  \end{pmatrix} = \ee^{ \ii \,\text{sign}(\epsilon)\phi }\dfrac{1}{\sqrt 2}\begin{pmatrix}0 \\ 1 \\ -\text{sign}(\epsilon) \ii  \end{pmatrix}
\end{equation}
where $\,\text{sign}(\epsilon)= \epsilon /|\epsilon |$. The singularity of $\Psi_+$  at $\infty$ can also be cured  using gauge freedom. We define $\Psi_+^B(k,\phi) := \ee^{-\ii \,\text{sign}(\epsilon)\phi} \Psi_+(k,\phi)$, implying
$
\lim_{k\rightarrow 0} \Psi_+^A(k,\phi) = \tfrac{1}{\sqrt 2}(0,1,-\,\text{sign}(\epsilon) \ii)
$
so that $\Psi_+^B$ is \emph{regular} on the punctured $(k_x,k_y)$ plane where the origin is removed $(\mathbb R\setminus\{0\}) \cup \{\infty\}$. The problem has been \emph{compactified} at $\infty$, in the sense that all the infinite directions are equivalent so that we can consider them as a limiting single point. Importantly, this is not possible without odd-viscosity ($\epsilon  = 0$). Indeed
\begin{equation}
\lim_{k\rightarrow \infty} \Psi_+(k,\phi) = \dfrac{1}{\sqrt 2}\begin{pmatrix} 1 \\ \cos\phi \\ \sin \phi  \end{pmatrix}
\end{equation}
so that the singularity is impossible to remove by the gauge freedom: the problem is \emph{not compactifiable} at $\infty$. This is illustrated in Figure \ref{fig:Compactornot}.

\begin{figure}
\centering
	\includegraphics[scale=0.78]{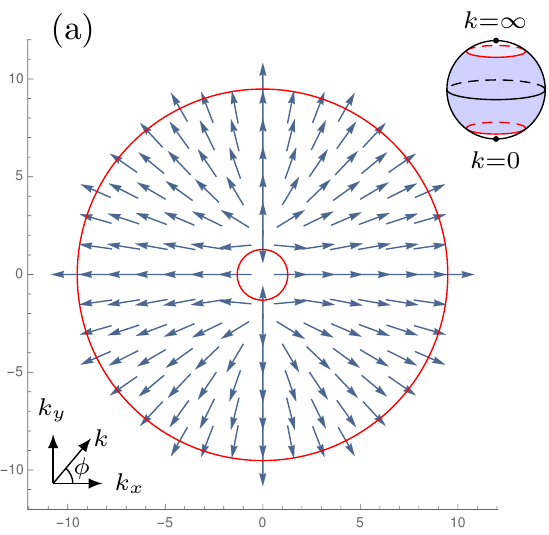}
	\includegraphics[scale=0.78]{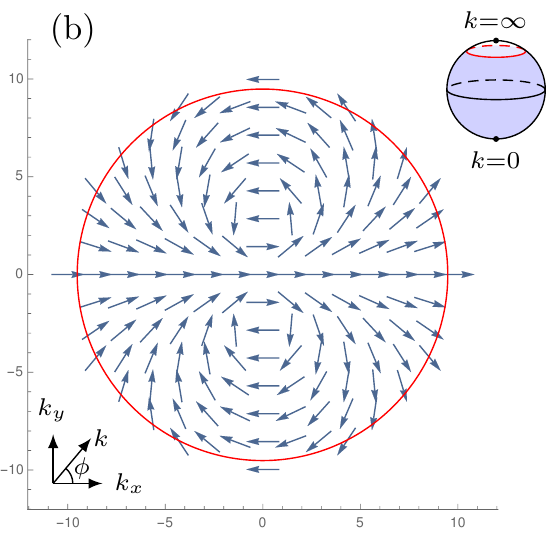}
	\includegraphics[scale=0.78]{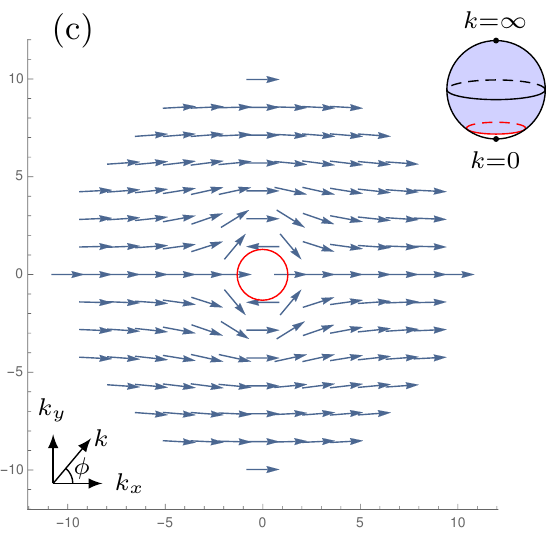}
	\caption{Normalized second component $\hat{u}_{+}/|\hat{u}_{+}| \in \mathbb C$ of positive frequency inertia-gravity eigenmodes $\Psi+$ for $f=1$ and $\epsilon  = 0.2$ in the $(k_x,k_y)$ plane, or equivalently on the sphere where the parallels correspond to circles of fixed radius $k$. (a) $\Psi_+$ is singular at $0$ and at $\infty$, but each singularity can be cured up to a phase: (b) at $0$ with $\Psi_+^A$ or (c) at $\infty$ with $\Psi_+^B$ . The problem is compactified in the sense that each singularity can be considered as a single point since it exists an eigenvector $\Psi$ regular (for all the components) in its neighborhood. A non-vanishing Chern number $C_+$ then captures the impossibility of finding a $\Psi$ regular everywhere. In the case where $\epsilon =0$ neither $\Psi_+^B$ nor (c) does exist so that the Chern number is not even well-defined.\label{fig:Compactzeroandinf}}
\end{figure}

So far the answer to question 1 is positive, but we defined two families $\Psi_+^A$ and $\Psi_+^B$ that differ by the choice of a phase, illustrated in Figure \ref{fig:Compactzeroandinf}. The answer to question 2 is captured by the Chern number. The latter is well defined as a topological invariant \emph{for compact manifolds only}. Here we identify $\mathbb R^2 \cup \{\infty\}$ with the two-sphere $S^2$ (see Figure \ref{fig:Compactzeroandinf}), for example through the stereographic projection, although we do not need any explicit transformation.  
The computation of the \textit{topological} Chern number is now very standard through the integral of the \textit{geometrical} Berry curvature \citep{nakahara2003geometry}. The Berry curvature is the curl of a Berry connection, that allows one to compare two adjacent normalized eigenmodes in parameter space. For $\alpha = A,\,B$ the Berry connection is $\mathbf A^\alpha_+ = -\ii \langle \psi_+^\alpha, \nabla \psi_+^\alpha \rangle$ and the Berry curvature is $\mathbf B_+ = \nabla \times \mathbf A^\alpha_+$, independent of $\alpha$. The connection depends on the phase of the eigenmodes, but the curvature is gauge independent. 

On $\mathbb R^2 \setminus \{0\}$ one has $\psi_+^B(k,\varphi)= \ee^{-\ii (\text{sign}(f)+\text{sign}(\epsilon))\phi} \psi_+^A(k,\varphi)$ and thus $\mathbf A_+^B = \mathbf A_+^A - (\text{sign}(f) + \text{sign}(\epsilon)) \nabla \phi $.
The Chern number is the flux of the Berry curvature through the whole (compactified) plane, namely
\begin{equation}
C_+ = \dfrac{1}{2\pi} \int_0^\infty \dd k \int_0^{2\pi} k \dd \phi \, \mathbf B_+ \cdot \mathbf e_z \ .
\end{equation}

Splitting $\mathbb R^2 \cup \{\infty\} = D_< \cup D_>$, respectively the disk $\{k\leq 1\}$ an its complementary, we apply Stokes theorem on each part and are left with a contribution at the border $k=1$. Explicitly
\begin{equation}
C_+ = \dfrac{1}{2\pi} \int_0^{2\pi}  \dd \phi \, (\mathbf A_+^A - \mathbf A_+^B) \cdot  \mathbf e_\phi= \text{sign}(f) + \text{sign}(\epsilon) \ .
\end{equation}

\cite{souslov2018topological} reached a similar conclusion for the bulk topology when computing the integral of the Berry curvature over the whole plane $(k_x,k_y)$, \textcolor{black}{and  this result can be recovered for the Dirac Hamiltonian in arbitrary dimension \cite{bal2018continuous}}.
%Our contribution here has been to prove that this quantity is a Chern number.
 Note that if we start with $\epsilon =0$ the latter derivation leads to $C_+ = \,\text{sign}(f)$, but this is \emph{not} %a topological invariant 
 \textcolor{black}{a Chern number}
 anymore, even if the Berry curvature integrated on the non-compact manifold $\mathbb R^2$ is finite.

\subsection{Topology of the lower and middle bands}

The Chern number of the lower band is by the symmetry of the system $\Psi_-(k_x,k_y,f,\epsilon ) = \Psi_+(-k_x,-k_y,-f,-\epsilon )$, leading immediately to
$C_- = - (\,\text{sign}(f) + \text{sign}(\epsilon)) = - C_+$. Finally the normalized family of eigenvectors associated to $\omega =0$ is 
\begin{equation}
\psi_0(k_x,k_y) = \dfrac{1}{\sqrt{k^2 +(f-\epsilon k^2)^2}} \begin{pmatrix}
f-\epsilon  k^2 \\ \ii k \sin \phi \\ -\ii k \cos \phi
\end{pmatrix} 
\end{equation}
In particular
$
\lim_{k\rightarrow 0} \Psi_0 = ( \,\text{sign}(f) , 0 , 0 )$ and $\lim_{k\rightarrow \infty} \Psi_0 = ( -\text{sign}(\epsilon), 0, 0 )   
$
so that $\Psi_0$ is regular  on the whole compactified plane $\mathbb R \cup \{\infty\}$: it is a global continuous section on a compact manifold. Thus
$
C_0 = 0
$.
Again, notice that when $\epsilon  = 0$ one has
$
\lim_{k\rightarrow \infty} \Psi_0 = ( 0, \ii \sin\phi ,-\ii \cos\phi )
$
so that the problem is not compactifiable at $\infty$.

\subsection{Summarizing}

If $f$ and $\epsilon $ have the same sign $s=\pm$, then the Chern number for the three wave bands are $C_+ = -C_- = 2s$ and $C_0 =0$. If $f$ and $\epsilon $ have opposite sign, then $C_+ = C_- = C_0 = 0$.   We now compute the wave spectrum when two hemispheres are glued together, with a given value of odd viscosity $\epsilon$. According to the bulk-interface correspondence, and whatever the sign of $\epsilon$, we expect that 2 unidirectional edge states should fill each frequency gaps between the flat band of geostrophic modes and the inertia-gravity wave bands.

\section{Interface: an equator between two flat hemispheres}

\begin{figure}
\centering
    \raisebox{1.3cm}{\includegraphics[scale=0.9]{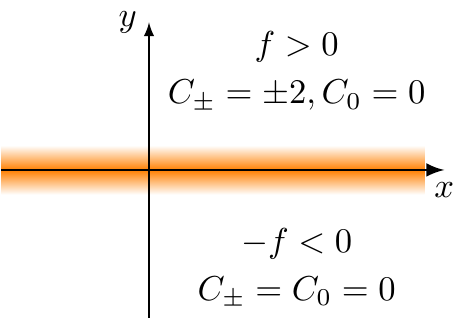}}
    \hspace{1cm}
	\includegraphics[scale=0.4]{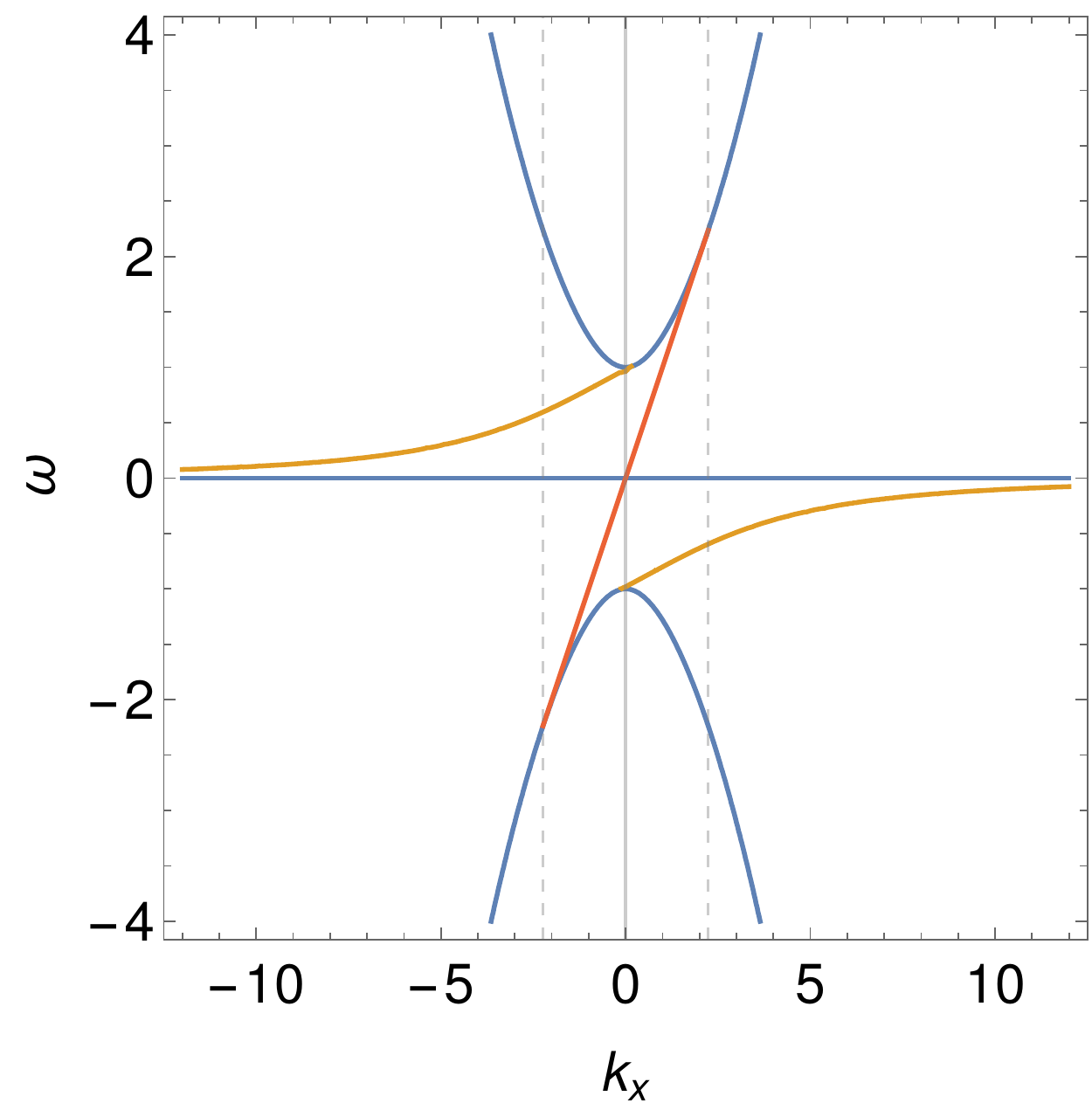}
	\caption{Left: sharp interface between two hemispheres interpreted as two distinct topological phases. Right: Dispersion relation of the modes confined at the interface (a Yanai-like wave in orange and a Kelvin-like wave in red) for $f=1$ and $\epsilon =0.2$: in each gap one has two topological solutions, in agreement with the bulk-edge correspondence. The black curve is the limit of the bulk inertia-gravity wave bands area. The vertical dashed lines correspond to $|k_x|=k_0$ where the Kelvin mode merges into the bulk by compactification. This cut-off tends to infinity when odd viscosity $\epsilon$ tends to zero.  \label{fig:interface_modes}}
\end{figure}

We consider a sharp interface at $y=0$ between two hemispheres that can now be interpreted as two distinct topological phases (see Figure \ref{fig:interface_modes}(left)). On the upper half-plane  $f$ is constant and positive, on the lower one $-f <0$. The parameter $\epsilon$ is constant and positive on the whole plane. To simplify the computations below we assume $f\epsilon <1/4$.

The translation invariance is preserved in the longitudinal direction, we look for normal modes $(\eta, u, v) = \ee^{\ii (\omega t -k_x x)}(\tilde \eta, \tilde u, \tilde v)$, the latter vector being a function of variable $y$ and parameters $k_x$ and $\omega$. Dropping the tilde, it is ruled by
\begin{align}
\ii \omega \eta &= \ii k_x u - \partial_y v \label{eq:first}\\
\ii \omega u &= \ii k_x \eta + (\pm f - \epsilon  k_x^2) v + \epsilon  \partial_{yy} v\\
\ii \omega v &= -\partial_y \eta - (\pm f - \epsilon  k_x^2) u - \epsilon  \partial_{yy} u  \, .
\end{align}
There is a redundancy in the system as $\eta$ is completely determined by $u$ and $v$ from \eqref{eq:first}. This can be eliminated, leading to a two dimensional problem of order two:
\begin{align}
\label{eq:interface_ode_v}\big(\epsilon  \partial_{yy}  - \dfrac{k_x}{\omega} \partial_y  + (\pm f - \epsilon  k_x^2)\big)v &= \dfrac{\ii}{\omega}(\omega^2-k_x^2) u \\
\label{eq:interface_ode_u}\big(\epsilon  \partial_{yy}  + \dfrac{k_x}{\omega} \partial_y  + (\pm f - \epsilon  k_x^2)\big)u &= -\dfrac{\ii}{\omega}(\partial_{yy}+\omega^2) v
\end{align}
At the interface we impose the continuity of $u$ and $v$ as well as their first derivative
\begin{equation}\label{eq:interface_gluing}
u|_{y=0^-} = u|_{y=0^+}, \quad v|_{y=0^-} = v|_{y=0^-}, \quad \partial_y u|_{y=0^-} = \partial_y u|_{y=0^+}, \quad \partial_y v|_{y=0^-} = \partial_y v|_{y=0^-}
\end{equation}
which implies the same for $\eta$. We look for the solutions that are localized near the interface, in the sense that they vanish away from it, when $y \rightarrow \pm \infty$. We first solve the problem in each half-plane by decomposing
\begin{equation}
u(y) = \left\lbrace \begin{array}{ll}
u_\uparrow(y), & y>0,\\
u_\downarrow(y), & y<0,\\
\end{array}\right.
\end{equation} and similarly for $v$, then glue the solutions at the interface through \eqref{eq:interface_gluing} and count the remaining degrees of freedom, leading to a number of interface modes. This number is algebraic, its sign being determined by $\partial \omega/\partial k_x$. The result is given in Figure \ref{fig:interface_modes} with $n=2$ modes in each gap. We now explain how to reconstruct it.

\subsection{The compactified Kelvin wave} It is easy to see that $u\equiv 0$ in the whole plane implies $v\equiv 0$. In this section we only assume $v \equiv 0$ and look for non-trivial $u$. By \eqref{eq:interface_ode_v} we infer $k_x^2 = \omega^2$, and $u$ is ruled by \eqref{eq:interface_ode_u}, a homogeneous equation of order 2. The solution depends on the relative sign of $k_x$ and $\omega$.

Case 1: $k_x = \omega$. On the upper half-plane the solution is of the form
	\begin{equation}\label{eq:q_up}
	u_\uparrow(y) = A_\uparrow \ee^{q_{\uparrow+} y} + B_\uparrow \ee^{q_{\uparrow-} y}
\quad
	\text{where}
\quad
	q_{\uparrow\pm} = - \dfrac{1}{2\epsilon } \big(\dfrac{k_x}{\omega} \pm \sqrt{1+4\epsilon (\epsilon k_x^2-f)}\big)
	\end{equation}
that is always well defined as long as $f\epsilon  \leq 1/4$. Notice that $q_{\uparrow\pm}$ (as well as $A_\uparrow$ and $B_\uparrow$) depends on $k_x$ and $\omega$. For $u_\uparrow$ to vanish at $y\rightarrow \infty$ we have either $q_{\uparrow\pm} <0$ or $A_\uparrow/B_\uparrow =0$. Here $q_{\uparrow+}<0$ for all $k_x$ and $q_{\uparrow-}<0$ only for $|k_x| < k_0 := \sqrt{f/\epsilon }$, so that 
\begin{equation}
u_\uparrow(y) = \left\lbrace \begin{array}{ll}
A_\uparrow \ee^{q_{\uparrow+} y} + B_\uparrow \ee^{q_{\uparrow-} y}, & |k_x| < k_0,\\
A_\uparrow \ee^{q_{\uparrow+} y}, & |k_x| \geq k_0.
\end{array}\right.
\end{equation}
In the lower half-plane, one has similarly
\begin{equation}\label{eq:q_down}
q_{\downarrow\pm} = - \dfrac{1}{2\epsilon } \big(\dfrac{k_x}{\omega} \pm \sqrt{1+4\epsilon (\epsilon k_x^2+f)}\big)
\end{equation}
but this time we select the positive roots, so that $u_\downarrow$ vanishes when $y \rightarrow -\infty$. Here $q_{\downarrow+} <0$ and $q_{\downarrow-} >0$ for all $k_x$ so that $u_\downarrow(y) = B_\downarrow \ee^{q_{\downarrow-} y}$. From the interface condition \eqref{eq:interface_gluing} we infer two relations between $A_\uparrow$, $B_\uparrow$ and $B_\downarrow$ for $|k_x| < k_0$ whereas $A_\uparrow = B_\downarrow = 0 $ for $|k_x| \geq k_0$. More precisely
\begin{equation}\label{u_kelvin}
u(y) = \left\lbrace \begin{array}{lll}
A_\uparrow \Big(\ee^{q_{\uparrow+} y} -\dfrac{q_{\uparrow+}-q_{\downarrow-}}{q_{\uparrow-}-q_{\downarrow-}} \ee^{q_{\uparrow-} y} \Big) & y>0, &|k_x|<k_0\\
A_\uparrow \dfrac{q_{\uparrow-}-q_{\uparrow+}}{q_{\uparrow-}-q_{\downarrow-}}\ee^{q_{\downarrow-} y} & y<0,& |k_x|<k_0\\
0 & |k_x|\geq k_0
\end{array}\right.
\end{equation}
We are left with one degree of freedom $A_\uparrow$, a positive dispersion relation $\omega=k_x$ that is moreover compactified: the solution exists only for $|k_x|<k_0$. Note that $k_0 \rightarrow \infty$ as $\epsilon  \rightarrow 0$. {\color{black} Moreover, in that limit $q_{\uparrow +}\sim -\tfrac{1}{\epsilon}$, $q_{\uparrow -} \rightarrow - f$ and $q_{\downarrow -} \rightarrow f$. For $y>0$, $\ee^{q_+ y} \rightarrow 0$ so that \eqref{u_kelvin} becomes, after renormalizing $A_\uparrow= \epsilon\,\tilde A_\uparrow $,
\begin{equation}
   u(y) \mathop{\longrightarrow}\limits_{\epsilon \rightarrow 0} \tilde A_\uparrow (-2f)^{-1} \, \ee^{-f|y|}, \qquad y  \in \mathbb R,\quad  k_x \in \mathbb R.
\end{equation}
It is remarkable that the limit $\epsilon \rightarrow 0 $ coincides with the classical (non-compactified) Kelvin wave solution obtained when $\epsilon=0$, in which case the order of the partial differential equation is lowered.}
%We recover the usual (non-compactified) Kelvin wave  exponentially trapped around the equator \citep{matsuno1966quasi}.

Case 2: $k_x = -\omega$. In that case $q_{\uparrow/\downarrow\pm}$ have the same expression but their sign change. A similar inspection leads to $u_\uparrow(y) = A_\uparrow \ee^{q_{\uparrow+}y}$ for $|k_x|\geq k_0$ and vanishes otherwise, and $u_\downarrow(y) = B_\downarrow \ee^{q_{\downarrow-}y}$ for $k_x \in \mathbb R$. The gluing condition implies $A_\uparrow = B_\downarrow = 0$, so that $u \equiv 0$.

\subsection{The compactified Yanai wave} 

In this section we assume $k_x^2 \neq \omega^2$ and $v\neq 0$. In that case $u$ is entirely fixed by $v$ through equation \eqref{eq:interface_ode_v}, and one can moreover combine \eqref{eq:interface_ode_v} and \eqref{eq:interface_ode_u} to get a fourth order homogeneous equation for $v$. In the upper half-plane it reads
\begin{equation}\label{eq:yanai_v_order4}
\Big(\epsilon ^2 \partial^{(4)}_{y}  + (2\epsilon ( f-\epsilon  k_x^2)-1) \partial^{(2)}_{y} + ( f-\epsilon  k_x^2))^2-(\omega^2-k_x^2)\Big)v = 0
\end{equation} 
The corresponding algebraic equation always admits real solutions as long as $f \epsilon  \leq 1/4$, given by $s^2 = S_\pm$ with
\begin{equation}\label{eq:def_Spm}
	S_\pm = \dfrac{1}{2\epsilon ^2} \Big( 1+2\epsilon (\epsilon  k_x^2- f) \pm \sqrt{1+4 \epsilon  (\epsilon  \omega^2-f)} \Big)
\end{equation}
In order to get a non trivial mode at the interface, we need both $S_+>0$ and $S_->0$. This implies
\begin{equation}\label{eq:interface_bulk_limit}
\Delta(k_x,\omega) := k_x^2-\omega^2 + (f-\nu k_x^2)^2 > 0
\end{equation}
In the region where $\Delta(k_x,\omega) \leq 0$ oscillatory solutions exist: they are the normal modes from the bulk which are still allowed in the interface setting. This region corresponds to the projection of the bulk bands $\omega_\pm(k_x,k_y)$ from \eqref{eq:omegapm} for all values of $k_y \in \mathbb R$.  Thus \eqref{eq:interface_bulk_limit} delimits the gaped region in the interface setting. In this region one has four real solution to \eqref{eq:yanai_v_order4}
\begin{equation}
s_{\uparrow1} = \sqrt{S_+}, \quad s_{\uparrow2} = \sqrt{S_-}, \quad s_{\uparrow3} = -\sqrt{S_+}, \quad s_{\uparrow4} = -\sqrt{S_-}
\end{equation}
and similarly for $s_{\downarrow i}$, $i=1,\ldots,4$, where we replace $f$ by $-f$ in the expression of \eqref{eq:def_Spm}. By construction $s_{\uparrow/\downarrow1/2} >0$ and $s_{\uparrow/\downarrow3/4} <0$ regardless of $k_x, \omega$ or $f$. Consequently,
\begin{equation}
v(y) = \left\lbrace \begin{array}{ll}
	V_{\uparrow 3} \ee^{s_{\uparrow 3} y} + V_{\uparrow 4} \ee^{s_{\uparrow 4} y}, & y>0\\
	V_{\downarrow 1} \ee^{s_{\downarrow 1} y} + V_{\downarrow 2} \ee^{s_{\downarrow 2} y}, & y<0.
\end{array}\right.
\end{equation}
Moreover $u_{\uparrow}(y) = \lambda_{\uparrow 3} V_{\uparrow 3} \ee^{s_{\uparrow 3} y} + \lambda_{\uparrow 4}V_{\uparrow 4} \ee^{s_{\uparrow 4} y}$ and $u_{\downarrow}(y) = \lambda_{\downarrow 1} V_{\downarrow 1} \ee^{s_{\downarrow 1} y} + \lambda_{\downarrow 2}V_{\downarrow 2} \ee^{s_{\downarrow 2} y}$ where 
\begin{equation}
\lambda_{\uparrow/\downarrow i} = \dfrac{\omega}{\ii (\omega^2-k_x^2)} \big( \epsilon  s_{\uparrow/\downarrow i}^2 - \dfrac{k_x}{\omega}s_{\uparrow/\downarrow i} \pm f - \epsilon  k_x^2 \big),
\end{equation}
inferred from \eqref{eq:interface_ode_v}. The four free parameters $ V_{\uparrow 3},  V_{\uparrow 4}, V_{\downarrow 1}$ and $V_{\downarrow 2}$ are constrained by the four gluing conditions \eqref{eq:interface_gluing}, so that a non-trivial solution exists only if $\det(M) =0$ where
\begin{equation}
M = \begin{pmatrix}
1 & 1& -1 &-1 \\ s_{\uparrow 3} & s_{\uparrow 4} & -s_{\downarrow 1}&-s_{\downarrow 2} \\
\lambda_{\uparrow 3} & \lambda_{\uparrow 4} & -\lambda_{\downarrow 1}&-\lambda_{\downarrow 2} \\
\lambda_{\uparrow 3}s_{\uparrow 3} & \lambda_{\uparrow 4}s_{\uparrow 4} & -\lambda_{\downarrow 1}s_{\downarrow 1}&-\lambda_{\downarrow 2}s_{\downarrow 2} 
\end{pmatrix} \, .
\end{equation} 
There is no simple expression for the dispersion relation $\omega = f(k_x)$ such that $\det(M)=0$, however the latter gives an implicit relation between $k_x$ and $\omega$, namely
\begin{equation}\label{eq:Yanai_implicit}
\epsilon ^2(s_{\downarrow 1}-s_{\uparrow 3})(s_{\downarrow 2}-s_{\uparrow 3})(s_{\downarrow 1}-s_{\uparrow 4})(s_{\downarrow 2}-s_{\uparrow 4}) + 2 f \dfrac{k_x}{\omega} (s_{\downarrow 1}+ s_{\downarrow 2}-s_{\uparrow 3} - s_{\uparrow 4}) - 4 f^2 = 0
\end{equation}
that can be computed numerically. This way we obtain the Yanai wave of Figure \ref{fig:interface_modes}: one in each gap. An expansion around $k_x \rightarrow \pm \infty$ shows the behavior $\omega \sim \mp\tfrac{ f}{2(1+k_x^2 \epsilon ^2)}$, so that the dispersion relation connects (asymptotically) the middle band $\omega=0$ to the upper band at $k_x=0$.  On the other hand for a finite value of $k_x$, an expansion around $\epsilon\rightarrow 0$ leads to $\omega \sim \mp f$: the compactified Yanai waves is an inertial wave in this limit, \textcolor{black}{as in \cite{iga1995transition} where $\epsilon=0$}.  For this mode the rank of $M$ is 3 so that we only have one free parameter among $V_{\uparrow 3},  V_{\uparrow 4}, V_{\downarrow 1}$ and $V_{\downarrow 2}$: we have  $n=1$ Yanai mode in each gap.

%Note that in principle there is also a branch $k_x=-\omega$ that is solution to \eqref{eq:Yanai_implicit} for $|k_x|\leq k_0$, but we remove it as it is forbidden by assumption. The careful reader might object that this branch is crossing the Yanai mode, so that we also remove this crossing point from the picture and are left with a discontinuity of the dispersion relation. This is nothing but an artefact of the way we solve the problem, we take care of this technical detail in supplementary material.
The last case $k_x^2= \omega^2$ with $v\neq0$ is rather tedious, but it can be checked that no extra mode appear in this case. This is done in supplementary material. 

\section{\textcolor{black}{Bulk-boundary correspondence and boundary conditions}}

\begin{figure}
\centering
\includegraphics[width=\textwidth]{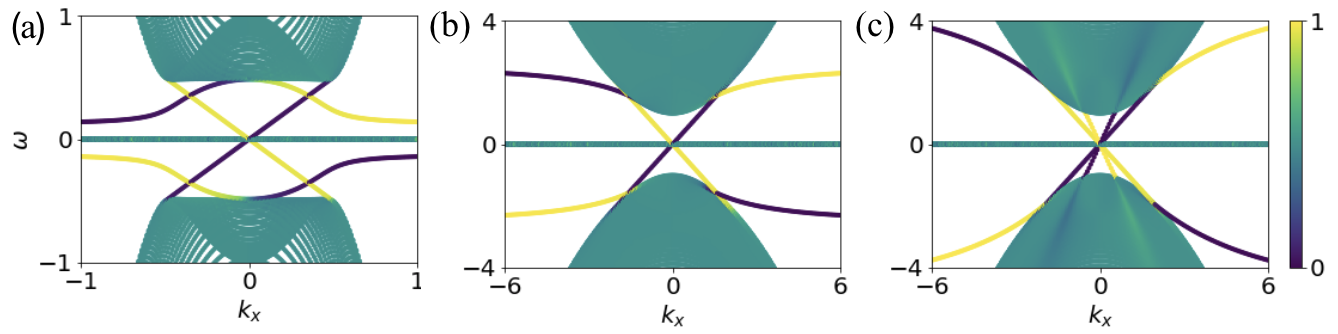}
	\caption{Dispersion relation of the shallow water model in a cylinder geometry with  $f=1$, $g=H=1$, using a numerical code \citep{dedalus}. a) no-slip boundary condition with large odd viscosity ($\epsilon=4$, $L_y=120$) b) no-slip  boundary condition with weak odd viscosity ($\epsilon=0.2$, $L_y=30$)  c) stress-free boundary condition with same parameters as in case b. The color code indicates localization of the modes in $y$ direction (from $0$ at $y=0$ to $1$ at $y=L_y$). \label{fig:spectrum_boundary}} 
\end{figure}

\textcolor{black}{
To emphasize the relevance of the interface studied above we briefly discuss the sharp boundary problem. Figure \ref{fig:spectrum_boundary} shows the spectrum of odd-shallow water waves computed numerically with \cite{dedalus} on a channel (or infinite strip) with sharp walls: $(x,y) \in \mathbb R \times [0,L]$, where $\epsilon$ and $f$ have the same sign. The boundary conditions are either: $v=0$, $u=0$ (no slip) or $v=0$, $\partial_xu-\partial_y v=0$ (stress-free).}

\textcolor{black}{Like for the interface, we recover the region of the (projected) bulk bands as well as edge modes in the gaped region, this time localized on each wall. In the following we focus on one of them (e.g $y=0$). In Figure \ref{fig:spectrum_boundary}(a) the number of modes crossing a fixed frequency line $\omega$ in the gap is 2, in agreement with the Chern number of the upper band \citep{souslov2018topological}. However this number becomes 1 either when $\omega_0$ is too close to the middle band $\omega_0=0$ or when $\epsilon$ is smaller: in that case the other mode never crosses the spectral gap windows $0<\omega<f$ (Figure \ref{fig:spectrum_boundary}(b)). Moreover, when changing the boundary condition, the total number of modes surprisingly jumps from 2 to 3 in Figure \ref{fig:spectrum_boundary}(c). }

\textcolor{black}{Thus it seems that the bulk-edge correspondence is not always satisfied for the sharp boundary problem since the number of edge modes depends on the choice of parameters and boundary conditions. On the other hand our interface setting with a canonical choice of gluing condition provides a remarkable case where the bulk-interface correspondence is fully satisfied: the number of modes localized at the interface inside the spectral region $0<\omega<f$ matches with the Chern number of the upper band.}

\section{Discussion and Conclusion} 

\textcolor{black}{The most representative difference of our study with the original description of equatorial waves by \cite{matsuno1966quasi} is that we consider a profile $f(y)$ with a sharp gap at the equator $(y = 0)$ and constant in each hemisphere, rather than linear variations of the Coriolis parameter with latitude ($f=\beta y$). In this celebrated equatorial beta plane configuration, all the modes are trapped along the equator, and described by Hermite functions \citep{matsuno1966quasi}. Here, in the case of a sharp interface, only the Yanai-like and Kelvin-like waves are exponentially trapped modes; the other modes are all delocalized. In addition, there is no Rossby wave. These properties are caused by the hypothesis of a constant value of $f$ in each hemisphere. If  we keep the hypothesis of a sharp equator but add back a constant gradient of Coriolis parameter into the problem, topological properties are preserved, but (i) the degeneracy of geostrophic modes is  lifted, with the emergence of Rossby waves (ii) all the modes are localized at the equator (like for Hermite functions), even if the Kelvin and Yanai waves remain more localized than the others, which can be understood by counting the zeros of eigenfunctions \citep{iga1995transition}. Using arguments based on the conservation of these zeros, \citep{iga1995transition} explained the global shape of equatorial wave spectra computed by Matsuno. In particular,  he found that the Yanai wave can be interpreted as an inertial wave. Our study brings a complementary point view, showing the topological origin of these properties, as the outcome of gluing two hemispheres with different topological indices. Our analysis also suggests that Yanai waves should be qualified as  mixed geostrophic-gravity waves rather than mixed Rossby-gravity waves, as they exist even in the absence of Rossby waves.  This result relied on the introduction of a regularization parameter, but we recover the spectrum of the original problem when the regularization coefficient tends to zero: there is no singular limit in the sharp interface case.}

To summarize, (i) it has been possible to assign a topological index to each hemisphere through the introduction of odd-viscous term, \textcolor{black}{as in \cite{souslov2018topological}. (ii) \citep{souslov2018topological} showed a range of parameters and boundary conditions for which the correspondence between the number of unidirectional edge states filling the frequency gap and the bulk Chern number were satisfied. Building on previous work by \cite{iga1995transition}, we noticed that the number of edge states that transit from one band to another in fact depends on the choice of the boundary condition, even in the presence of odd-viscosity. This new observation raises the question of the existence of a  bulk-boundary correspondence for fluids in particular, and for continuous media in general. (iii) We have considered the simpler case of a sharp interface separating the two hemispheres. This interface case bypasses the need to discuss boundary conditions, as the natural physical choice is to impose continuity of the fields and their derivatives. We have found in that case explicit analytical solutions exhibiting exactly two unidirectional modes in each gap confined along the equator, accordingly with the bulk-interface correspondence. Remarkably, this simple case provides a  solvable example of a bulk-interface correspondence in fluids, in a configuration where the interface in infinitely sharp. This paves the way towards an understanding of the bulk-edge correspondence in continuous media with boundaries. We will explain in a companion paper that the apparent paradox between the number of edge states and the bulk Chern number can indeed be explained, beyond the particular case of the shallow water model.}

\begin{acknowledgments}
C. T. is grateful to Gian Michele Graf and Hansueli Jud for many insightful discussions. P. D. and A.V. were partly funded by  ANR-18-CE30-0002-01 during this work, and thank L.-A. Couston for help with Dedalus. 
\end{acknowledgments}

\bibliographystyle{jfm}

\bibliography{jfm-topological}
%\newpage 
%\appendix 
\section*{Supplementary material: checking the possible remaining modes at the interface  \label{App:interface_lastcase}}

The last case to study is $v \neq 0$ and $k_x^2 = \omega^2$, equations \eqref{eq:interface_ode_v} and \eqref{eq:interface_ode_u} become
\begin{align}
\big(\epsilon  \partial_{yy}  - \dfrac{k_x}{\omega} \partial_y  + (\pm f - \epsilon  k_x^2)\big)v &= 0 \\
\big(\epsilon  \partial_{yy}  + \dfrac{k_x}{\omega} \partial_y  + (\pm f - \epsilon  k_x^2)\big)u &= -\dfrac{\ii}{\omega}(\partial_{yy}+\omega^2) v
\end{align}
from which we infer
\begin{align}
v_{\uparrow/\downarrow}(y) &= A_{\uparrow/\downarrow}\ee^{r_{\uparrow/\downarrow+}y} +  B_{\uparrow/\downarrow}\ee^{r_{\uparrow/\downarrow-}y} \\
u_{\uparrow/\downarrow}(y) &= \alpha_{\uparrow/\downarrow+}A_{\uparrow/\downarrow}\ee^{r_{\uparrow/\downarrow+}y} +  \alpha_{\uparrow/\downarrow-}B_{\uparrow/\downarrow}\ee^{r_{\uparrow/\downarrow-}y} + C_{\uparrow/\downarrow}\ee^{q_{\uparrow/\downarrow+}y} +  D_{\uparrow/\downarrow}\ee^{q_{\uparrow/\downarrow-}y}
\end{align}
where
\begin{equation}
r_{\uparrow/\downarrow} = \dfrac{1}{2\epsilon } \big(\dfrac{k_x}{\omega} \pm \sqrt{1+4\epsilon (\epsilon k_x^2-\pm f)}\big)
\end{equation}
with the last $\pm$ sign on the right hand side refers to $\uparrow/\downarrow$. The roots $q_{\uparrow/\downarrow\pm}$ are the one from \eqref{eq:q_up} and \eqref{eq:q_down} and furthermore
\begin{equation}
\alpha_{\uparrow/\downarrow\pm} = - \dfrac{i}{\omega}(r_{\uparrow/\downarrow\pm}^2 + \omega^2) \big(\epsilon  r_{\uparrow/\downarrow\pm}^2 + \dfrac{k_x}{\omega}r_{\uparrow/\downarrow\pm} + \pm f -\epsilon  k_x^2\big)^{-1}
\end{equation}
where $\pm f$ refers to $\uparrow/\downarrow$.

Case 1: $k_x=\omega$. One has $r_{\uparrow+}\geq0$ for all $k_x$ and $r_{\uparrow-}\geq0$ for $|k_x|\leq k_0 = \sqrt{f/\nu}$, so that $v_{\uparrow}(y) = B_\uparrow \ee^{r_{\uparrow-}y}$ for $|k_x|> k_0$ and vanishes otherwise. On the lower half-plane $r_{\downarrow+}>0$  and $r_{\downarrow-}<0$ for all $k_x$ so that $v_{\downarrow}(y) = A_\downarrow \ee^{r_{\downarrow+}y}$. The gluing condition \eqref{eq:interface_gluing} implies $B_\uparrow = A_\downarrow = 0$ so that $v\equiv 0$, which is forbidden by assumption (equivalently we are back to the Kelvin solution). There is no extra interface mode in that case.

Case 2: $k_x=-\omega$. One has $r_{\uparrow-} <0$ for all $k_x$ and $r_{\uparrow+}<0$ for $|k_x|<k_0$ whereas $r_{\downarrow-} <0$ and $r_{\downarrow+} >0$ for all $k_x$, so that
\begin{align}
v_{\uparrow}(y) &= A_{\uparrow} \ee^{r_{\uparrow+}y} + B_{\uparrow} \ee^{r_{\uparrow-}y}\\
v_{\downarrow}(y) &= A_{\downarrow} \ee^{r_{\downarrow+}y} 
\end{align}
with $A_\uparrow = 0$ for $|k_x| \geq k_0$. Similarly 
\begin{align}
u_{\uparrow}(y) &= \alpha_{\uparrow+} A_{\uparrow} \ee^{r_{\uparrow+}y} + \alpha_{\uparrow-}B_{\uparrow} \ee^{r_{\uparrow-}y} + C_{\uparrow} \ee^{q_{\uparrow+y}}\\
u_{\downarrow}(y) &= \alpha_{\downarrow+}A_{\downarrow} \ee^{r_{\downarrow+}y} + D_{\downarrow} \ee^{q_{\downarrow-y}}
\end{align}
with $C_{\uparrow} = 0$ for $|k_x|\leq k_0$ since $q_{\uparrow+}\geq0$ in that case, see \eqref{eq:q_up}. For $|k_x|\geq k_0$ the gluing condition \eqref{eq:interface_gluing} implies $B_\uparrow = A_\downarrow =0$ so that $v \equiv 0$ which is forbidden by assumption. For $|k_x|<k_0$ the four free parameter $A_\uparrow, B_\uparrow, A_\downarrow$ and $D_\downarrow$ are constrained by the four gluing conditions \eqref{eq:interface_gluing}. There exist a non-trivial solution only if $\det N(k_x,-k_x) =0$ with
\begin{equation}
N= \begin{pmatrix}
1 & 1& -1 &0 \\ r_{\uparrow +} & r_{\uparrow -} & -s_{\downarrow +}&0 \\
\alpha_{\uparrow +} & \alpha_{\uparrow -} & -\alpha_{\downarrow +}& -1 \\
\alpha_{\uparrow +}r_{\uparrow +} & \alpha_{\uparrow -}r_{\uparrow -} & -\alpha_{\downarrow +}s_{\downarrow +}&-q_{\downarrow -} 
\end{pmatrix}
\end{equation}
One can check numerically that along $(k_x,-k_x)$ this occurs only twice, and exactly at the crossing with the Yanai wave dispersion relation. This ensures the continuity of the latter and confirms that there is no other mode in that case.

\end{document}